\documentclass[sigconf,screen]{acmart}

\usepackage{booktabs}
\usepackage{tabularx}
 
\AtBeginDocument{%
  }

\setcopyright{acmlicensed}
\copyrightyear{2026}
\acmYear{2026}
\acmDOI{XXXXXXX.XXXXXXX}
\acmConference[EASE 2026]{The 30th International Conference on Evaluation and Assessment in Software Engineering}{9–12 June, 2026}{Glasgow, Scotland, United Kingdom}

\acmISBN{978-1-4503-XXXX-X/2018/06}

\begin{document}

\title{Comprehension Debt in GenAI-Assisted Software Engineering Projects}

\author{Muhammad Ovais Ahmad} 
\email{ovais.ahmad@kau.se}
\orcid{0000-0002-7885-0369}
\affiliation{
  \institution{Department of Computer Science, Karlstad University}
  \city{Karlstad}
  \country{Sweden}
}

\renewcommand{\shortauthors}{Ahmad, M.O. }

\begin{abstract}
Generative Artificial Intelligence (GenAI) tools (e.g., ChatGPT, Calude) have rapidly become integral to software development. These tools are especially attractive to students, as they can reduce cognitive load. However, their adoption also introduces a socio-cognitive risk: the accumulation of Comprehension Debt (CD). CD refers to the growing gap between what a development team knows about its codebase and what it actually needs to understand in order to maintain and modify it effectively.
This qualitative study investigate how GenAI tools contribute to CD in the context of an undergraduate software engineering project. Our study is based on 621 reflective diaries from 207 students over eight weeks. We identify four CD accumulation patterns and one mitigating pattern in students' use of GenAI tools. The four accumulation patterns include: (1) AI-as-black-box code acceptance, (2) context-mismatch debt, (3) dependency-induced atrophy, and (4) verification-bypass. In contrast, the mitigating pattern involves students using GenAI as a comprehension scaffold, allowing them to build a deeper understanding of the code. We argue that CD is distinct from traditional technical debt because it resides in the collective cognition of development teams rather than in the codebase itself. Our findings highlight the need for explicit pedagogical strategies to mitigate CD in software engineering education, emphasizing verification practices, structured retrospectives, and active learning assessments.  
\end{abstract}



\maketitle

\section{Introduction}
Generative AI (GenAI) coding assistants have transformed software development landscape. GenAI tools are now widely used by students and developers to explain errors, generate code snippets and so on.  They enhances code-authoring performance and improves learning experience for novice programmers  \cite{kazemitabaar2023studying}. Human-AI collaboration boosts intrinsic motivation and reduces programming anxiety compared to human pairing \cite{fan2025impact}. AI impact on programming education will fundamentally change how programming is taught in engineering \cite{qadir2023engineering, silva2024chatgpt}. Students using ChatGPT in introductory programming could submit correct solutions without acquiring underlying skills the assignments were designed \cite{leinonen2023using}. In university software engineering projects students develop programming skills and learn agile way of working in a collaborative environment. GenAI tools are particularly appealing to students because they can reduce cognitive load of unfamiliar tasks. However, GenAI tools introduce a structural risk. When students accepts GenAI code without understanding why it works, they externalized and deferred cognitive load. This deferred understanding accumulates as Comprehension Debt (CD). ``\textit{When teams produce code faster than they can understand it, it creates CD}'' \cite{Gorman2025}. It is cumulative gap between the demands a codebase makes on its developers and collective understanding of those developers possess. 

Unlike technical debt (TD), which resides in code artifacts and architectural decisions, CD resides in team cognition and shared mental models. It is therefore socio-cognitive rather than purely technical. CD cannot be detected by static analysis tools or automated tests, because it concerns what developers do not know about the system they are modifying. CD is also distinct from knowledge silos, as it describe uneven knowledge distribution. Whereas CD describes insufficient knowledge relative to system demands, regardless of distribution. 
This paper investigates CD in context of GenAI-assisted student agile projects. 

\textbf{Research Question}. \textit{In what ways do GenAI coding assistants contribute to, or mitigate, CD in software engineering projects?}

The paper is structured as follows. Section 2 presents background. Section 3 describes research method. Section 4 presents findings. Section 5 discusses findings, their implications, study limitation and future research directions. Section 6 concludes the paper.

\section{Background}
TD introduced by Cunningham \cite{cunningham1992wycash} and systematically extended by Kruchten et al. \cite{kruchten2012technical}, describes accumulated cost of expediency driven decisions in software development. A growing taxonomy identifies code debt, design debt, test debt, documentation debt, process debt and social debt among others \cite{ahmad2024pandora, ahmad2025technical}. In software engineering, CD refers to the future time cost of having to understand code before it can be safely changed, especially when code is produced faster than developers can realistically review and internalize it. The CD term was introduced by Jason Gorman, arguing that unchecked, rapidly generated code can accumulate a growing ``debt'' of understanding that must eventually be ``paid'' when teams need to debug, modify, or refactor the system \cite{Gorman2025}. Conceptually, CD extends TD and non-technical debt metaphors \cite{ahmad2024pandora, kruchten2012technical}. As GenAI tools increase output velocity, Gorman \cite{Gorman2025} argues that limiting factor shifts from writing code to understanding it, making CD a distinct and increasingly salient risk for maintainability and changeability in modern codebases. CD differs from TD in three key ways: (1) TD resides in artifacts; CD resides in cognition; (2) TD can be inspected; CD remains invisible until cognitive demand exceeds understanding; (3) TD accrues through design shortcuts; CD accrues when understanding lags behind code growth. 

This distinction aligns CD with socio-technical and cognitive theories \cite{sweller1994cognitive}, in which shared mental models are central to team performance. Cognitive load theory distinguishes intrinsic load (i.e., inherent task complexity), extraneous load (i.e, complexity added by poor task design), and germane load (i.e, cognitive effort that builds lasting mental schemas). GenAI tools may reduce intrinsic load by handling syntactic complexity and reduce extraneous load by clarifying documentation. However, when used in solution-substitution mode, they may reduce germane load which results into CD over time. Empirical studies reported a consistent tension between productivity and comprehension. Vaithilingam et al. \cite{vaithilingam2022expectation} found that GitHub Copilot users produced working code faster but understood generated code less than developers who wrote code manually. Hou et al. \cite{hou2024large} identified code understanding and trust calibration as among the most significant open challenges. In educational contexts, these dynamics are amplified \cite{leinonen2023using, finnie2022robots}. Students using ChatGPT and submit correct solution to introductory programming tasks without acquiring underlying skills \cite{leinonen2023using} . Liffiton et al. \cite{liffiton2023codehelp} introduced CodeHelp, an LLM-based tutoring tool that answers programming questions without providing direct solutions, motivated precisely by the concern that direct solution provision bypasses comprehension.  

\section{Methodology}
The data were collected in a software engineering project at Karlstad University, Sweden. The course follows a Scrum-based structure across four two-week sprints. Four to six students work in a team. The project encompassed seven functional requirements including user authentication, mood tracking with visualization, step counting, activity logging, leaderboard, accessibility features, and user profile system. Students were free to select their technology stack within broad constraints. Majority of teams in the course adopted Android Studio with Kotlin or Java, Firebase as backend, and GitLab for version control. No restrictions were placed on GenAI tool use. 

The most commonly reported tools were ChatGPT, Google Gemini (integrated into Android Studio), and GitHub Copilot (used by a few of students with active licenses). Students were assessed on technical deliverables, process adherence, and structured report submitted at course end. The primary dataset consists of 621 reflective diary entries written by 207 undergraduate students over the eight-week project. Diaries were written in English and reflect students’ individual experiences. All student learning diaries were anonymized. The diaries were not counted towards grading in the course. 
We conducted thematic analysis following six-phase approach of Braun and Clarke \cite{braun2006using}: corpus familiarization, initial open coding of all AI-related diary passages, pattern identification, pattern review against the full corpus, pattern definition and naming, and report production.  

\section{Results}

\subsection{Pattern 1: Black-Box Code Acceptance}
The most direct form of CD occurred when students accepted and committed AI-generated code without understanding of why it worked. In this pattern, AI tool functions as a solution provider. Code moves from AI response into codebase without mental-model construction that would enable future reasoning, modification, or debugging. For example, students routinely described a workflow of pasting complete XML layout files into ChatGPT, describing a desired visual arrangement in natural language, and receiving a modified file that they then used. The attraction of this approach was speed; the cost was comprehension. A student articulated that even when accepting generated code, integration itself needs some sort of understanding.  ``\textit{I often puzzle AI code into codebase. If I dont understand what chatgpt had given me it would still be hard to integrate it.}'' This shows that AI generates code may be locally correct, but inserting it into a larger codebase requires comprehension to locate integration point, understand data types involved, and recognize naming conflicts. Students who lacked even this threshold-level understanding reported integration failures.

The temporal structure of this pattern is significant because when code is accepted in one sprint but requires modification in next sprints. A student described their experiences that code initially worked well but broke down when changes were needed: ``\textit{Something that also hit me when I wanted to make changes to the code was that I had a really hard time to even understand simplest syntax of language}''. Here, CD incurred by accepting AI-generated code without engagement with syntax became blocking when student needed to exercise authorial control over code. A feature was delivered but the students didnt build their understanding and that debt became due at the moment of required modification. Another student explicitly recognized risk and drew a normative conclusion about conditions under which black-box acceptance is acceptable. `\textit{`As long as I understand the code I don’t see the problem with it, it’s just a time saver.}'' This conditional framing implies awareness that acceptance without understanding is problematic, even if  explicit reasoning was instrumental. 

\begin{table}[h]
\centering
\caption{Summary of CD patterns}
\begin{tabular} {p{0.9cm} p{2cm} p{1.5cm} p{2.9cm}}
\hline
\textbf{Pattern} & \textbf{Label} & \textbf{CD Effect} & \textbf{Core Mechanism} \\
\hline

P1 & Black-Box Code Acceptance & Accumulates CD & GenAI code integrated without understanding   \\

P2 & Context-Mismatch Debt & Accumulates CD & AI unaware of team codebase; outputs require heavy rework   \\

P3 & Dependency-Induced Atrophy & Accumulates CD & AI availability reduces efforts in independent comprehension   \\
 
P4 & Verification Bypass & Accumulates CD & AI inaccuracy undetected due to knowledge gaps   \\
 
\hline

\end{tabular}


\end{table}

\subsection{Pattern 2: Context-Mismatch CD}
A structurally distinct CD pattern emerged when GenAi tools are used standalone with no access to actual codebase, naming conventions and architectural decisions. Students give and ask GenAI tools isolated code fragments (e.g., a function, an error message, a class) and receive suggestions appropriate to fragment in isolation. Such code required adaptation and comprehension which the student often did not have it. A student reflected on this dynamic as: ``\textit{I still had to tweak and debug queries, since ChatGPT could not fully understand how database was built up, but it gave me a starting point}''. Another student framed context mismatch as a source of time cost rather than time saving: ``\textit{ChatGPT sometimes provided solutions that didn’t fully meet our project requirements or suggested code that was overly complex... it could take more time to adjust the code}''.... ``\textit{At times ChatGPT gave me overly complex answers that didn’t work in practice.}.'' Context mismatch was partially mitigated by codebase-aware tools. Students who integrated GenAI tools directly into their project-level codebase access produced more contextually appropriate suggestions: ``\textit{GPT4 was smarter and gave better advice in general, but Gemini was integrated into Android Studio and had access to the project itself so it could give feedback with better insight}''. However, Gemini contextual awareness was having challenges: ``\textit{Gemini however, while being proficient in giving advice relating to firebase, is really bad at debugging code}''.

\subsection{Pattern 3: Dependency-Induced Comprehension Atrophy}
Students’ sustained reliance on GenAI tools progressively reduced their time and efforts in independent comprehension (e.g., reading documentation, working through errors manually, tracing logic) that would otherwise have built robust mental models of codebase. The consequence was not necessarily incorrect code, but shallow understanding that left students unprepared for later, more complex tasks. 
`\textit{I did use ChatGPT for some general code structure in the beginning of project as I was in unfamiliar territory of React Native. I also used it to try to find solutions that I could not find any help on StackOverflow. It helped me get into how to work with React Native which is good but become a crutch that I tried my hardest to avoid}''. Students also described dependency risk in terms of its effect on problem-solving. ``\textit{It is very tempting to turn to ChatGPT for quick guidance, which sometimes lead to laziness in problem-solving}''. GenAI tools are beneficial at the beginning of learning trajectory, but continued reliance beyond that initial phase suppresses natural progression to deeper and self-directed comprehension.  

\subsection{Pattern 4: Verification Bypass} 
GenAI tools frequently generated incorrect, incomplete, or contextually inappropriate outputs. However, effective verification of AI outputs requires domain knowledge.
Students lacking sufficient understanding are unable to detect AI inaccuracies, which has compounding on CD. CD arose not from inaccuracy itself, but from students’ inability to detect it. Most students uses GenAI but lack of domain knowledge to identify error in complex code. Students described a range of inaccuracy modes, from subtly wrong code to confidently stated incorrect facts. ``\textit{....be careful, because AI give completely wrong information often, so one has to be vigilant and always double check}''.... ``\textit{The negatives about using chatgpt for code implementation is that sometimes it was completely wrong}''.  The verification bypass problem was recognized by multiple students who developed explicit verification practices as a response: ``\textit{ .... these tools can occasionally generate inaccurate or misleading responses, verifying their suggestion is crucial to avoid implementing incorrect solutions}''..... `` \textit{I often needed to adjust its [chatgpt] suggestions in order for it to be useful for project}''. However, not all students developed or applied consistent verification practices. GenAI code that appear to work was committed without cross-checking, and errors only surfaced later—at runtime. 
`\textit{ ... ChatGPT ... provided a lengthy solution that wasn’t useful. In the end, I solved issue with assistance of my team members}''. When verification is insufficient, inaccurate or erroneous AI code creep into codebase. Such erroneous code persists until runtime consequences force its discovery. The cost of discovery at that later point is substantially higher than the cost of verification at point of acceptance.

\subsection{GenAI as CD Mitigation}
Students used GenAI tools to improve their comprehension. In this mode, interactions focus on learning rather than mere code generation. Students use GenAI to explain code logic, clarify difficult concepts, and explore API functions. This behavior prioritizes active engagement over passive generation of code. 
``\textit{Most of the time I used it [chatgpt] to identify problems in my code and explanation about things I did not understand}''. 
A particularly important adaptive practice was explicit imposition of a comprehension gate before committing GenAI code. A student described this self-regulation practice as ``\textit{Some bits of code I got from AI but then I rewrote it and made sure I understood it before I used it. I didn’t just take generated code and put it to use and not understanding it}''. This rewrite before commit practice is a behavioral intervention that converts GenAI code generation into a comprehension building exercise. CD that would have accumulated through black-box acceptance was never incurred because students retained authorial and cognitive ownership of code. GenAI expand the scope of what a developer knows to know, showing existence of functions and patterns that would otherwise be invisible. ``\textit{... using ChatGPT for coding commands helped me discover built-in functions that I wasn’t aware of, which saved time by preventing the need to write completely new code from scratch}''.

\section{Discussion}
This paper investigated how GenAI coding assistants contribute to and mitigate CD in student agile software engineering projects. We identified four CD accumulation patterns and one CD mitigating pattern. Our findings suggest that GenAI tools function as amplifiers of epistemic orientation rather than as inherently beneficial or harmful technologies. This aligns with Vaithilingam et al \cite{vaithilingam2022expectation} findings that tools like GitHub Copilot allow developers to complete tasks faster but they often result in a lower rate of code understanding compared to manual code writing. The same tool that accelerates learning for one student may accumulate CD for another, depending on how it is cognitively and behaviorally integrated into software development practice.

We propose a conceptual model that explains CD accumulation and mitigation through the lens of cognitive load theory and epistemic orientation. Our findings shows two contrasting orientations (acceleration orientation and exploration orientation). In acceleration orientation, students primarily use GenAI as productivity enhancer and solution provider. GenAI outputs are accepted with minimal interrogation, reducing investment in germane cognitive load which is required to construct mental schema. Sweller \cite{sweller1994cognitive} highlighted that when external tools perform ``heavy lifting'' of schema construction, the learner's germane load is critically diminished. Germane load is important because it is the effort dedicated to processing and automating new information in students. While intrinsic and extraneous load may be temporarily reduced (e.g., faster syntax production, quicker debugging), reduced germane load leads to shallow mental models. Over time, this imbalance results in CD when codebase grows in functionality faster than team understanding grows in depth. In contrast, exploration orientation positions GenAI as comprehension scaffold. Students request explanations, rewrite AI-generated code, verify outputs against documentation, and use tool to discover conceptual relationships. In this mode, GenAI supports germane load rather than replacing it. Schema construction is strengthened, and CD is mitigated. 

Effective verification of GenAI outputs requires domain knowledge and conceptual understanding. However, students often turn to GenAI precisely because they lack such knowledge. Hou et al. \cite{hou2024large} identified this ``trust calibration'' as a significant open challenge in software engineering. Developers struggle to bridge gap between perceived correctness of AI suggestions and their actual logical validity \cite{hou2024large}. This creates a dangerous feedback loop. Meaning that lack of expertise leads to over-reliance on AI, which in turn prevents development of very expertise needed to audit the AI output. As a result, students who need to verify AI outputs are least equipped to do so. Patterns 1–3 increase likelihood of Pattern 4. Students who rely heavily on GenAI reduce opportunities to build domain knowledge, which in turn reduces their capacity to detect AI inaccuracies. Errors then appear in later stages (such as integration, runtime failure), which become a costly correction. This paradox has important pedagogical implications. For example, instructing students to ``verify AI outputs'' is not sufficient. Teachers needs to design tasks that builds knowledge which makes verification possible. Verification is a competence, not merely a behavior.

\subsection{CD, TD \& Socio-Technical Theory}
This study extends TD discourse by introducing CD as a socio-cognitive complement. Table 2 presents a structured comparison between acceleration and exploration orientations across cognitive, behavioral, and outcome dimensions. TD resides in code artifacts; CD resides in shared mental models. While traditional TD focuses on structural ``code smells,'' CD focuses on knowledge smells that occur when the developer's mental model becomes decoupled from code's logic. TD is measurable through structural indicators; CD becomes visible only when cognitive demand exceeds available understanding. CD can accumulate even when code quality is high and tests pass. A codebase may be technically sound while being cognitively opaque to its developers. This decoupling of artifact quality and team comprehension is particularly salient in AI-assisted development contexts.

 \begin{table}[t]
\caption{Comparison of GenAI epistemic orientations \& CD.}
\label{tab:orientation-model}
\centering
\small
\setlength{\tabcolsep}{4pt}
\renewcommand{\arraystretch}{1.2}
\begin{tabular}{p{0.18\linewidth} p{0.38\linewidth} p{0.37\linewidth}}
\hline
\textbf{Dimension} & \textbf{Acceleration Orientation} & \textbf{Exploration Orientation} \\
\hline
Primary goal & Speed / output delivery & Understanding / learning \\
AI role & Solution provider & Comprehension scaffold \\
Germane cognitive load & Reduced investment (schema construction deferred) & Supported / increased investment (schema construction promoted) \\
Typical behaviors & Copy--paste, minimal interrogation, prompt for finished code & Rewrite, request explanations, stepwise prompting, verify against docs/tests \\
Verification capacity & Often low (knowledge gaps limit error detection) & Higher (active engagement builds verification competence) \\
Expected outcome & CD accumulation (understanding lags behind code growth) & CD mitigation (understanding grows with code) \\
\hline
\end{tabular}
\end{table}

\subsection{Implications for Software Engineering Education}
Our findings suggest four concrete recommendations for software engineering educators. \textbf{First}, AI literacy instruction should be made explicit and practice-specific. Generic warnings about GenAI inaccuracy are insufficient. Drawing from cognitive apprenticeship model \cite{collins2018cognitive}, educators should ``model'' internal thought process of reviewing AI code, making invisible cognitive work of verification visible to students. Students benefit from structured guidance distinguishing scaffold-oriented use from bypass-oriented use. Providing concrete examples of rewrite-before-commit, explanation-first prompting, and documentation verification can operationalize productive AI engagement. 
\textbf{Second}, sprint retrospectives should include a structured comprehension surfacing component. Explicitly asking teams to identify parts of the codebase that are insufficiently understood makes CD visible and actionable. 
This practice aligns with social constructivism \cite{vygotsky1978mind, ahmad2024deep, ahmad2025strengthening} meaning that by discussing GenAI logic, students move code from ``black-box'' into their collective zone of proximal development. Identified gaps can then be addressed through targeted pair programming, code walkthroughs, or documentation requirements before next sprint. 
\textbf{Third,} assessment design must align incentives with understanding. When evaluation rewards only feature delivery, acceleration orientation becomes rational. ``Active Learning'' \cite{prince2004does} assessments should shift from evaluating the artifact to evaluating process of its creation. Incorporating oral code walkthroughs, explanation-based grading components, or re-implementation exercises shifts incentives toward exploration orientation. Assessments that include code explanation component requiring students to explain, in writing or orally, the logic of specific implemented features,  penalize the bypass modes and reward Pattern 5. 
\textbf{Fourth,} GenAI tool selection should account for context-awareness.  Codebase-integrated tools may reduce context-mismatch issues but do not eliminate the need for epistemic scaffolding. Tool sophistication does not substitute for comprehension building. Educators must ensure that higher tool accuracy does not lead to automation bias. Otherwise students will decreases their vigilance because GenAI tool appears competent.

\subsection{Limitations and Future Work}
This study is based on self-reported reflective diaries within a single software engineering course at Karlstad University, which may limit generalizability. However, the course context reflects a realistic agile development environment with authentic team collaboration and unrestricted GenAI tool usage. Therefore, the findings are intended to support analytical generalization to similar educational and early-stage professional contexts rather than broad population-level inference. We relies on self-reported reflective diaries, which may introduce biases such as selective recall or social desirability. To mitigate this, diaries were not graded, reducing incentives for strategic reporting, and were collected longitudinally across multiple sprints. Nevertheless, future work should triangulate these findings with behavioral data such as repository activity, IDE interaction logs, or controlled comprehension assessments. Future work should develop operational and measurable indicators to enable empirical validation. Potential indicators include: explanation latency, modification difficulty. Due to ethical and privacy constraints, raw diary data cannot be publicly shared. However, representative excerpts and descriptions of the analytical process are provided to support transparency and interpretability.

\section{Conclusion}

This paper examined how GenAI tools influence understanding in student software engineering projects. We introduced CD as a socio-cognitive construct describing gap between codebase demands and collective team understanding. We identified four CD accumulating patterns (black-box acceptance, context-mismatch, dependency atrophy, and verification bypass) and one CD mitigating pattern (AI as comprehension scaffold). We also articulate conceptual model of linking epistemic orientation, germane cognitive load investment, and CD accumulation. GenAI tools do not inherently undermine or enhance learning. Our study shows that they act as amplifiers of students existing orientation toward acceleration or exploration. Additionally, shows that verification competence must be intentionally cultivated through courses or module design. Otherwise students may find themselves in competence trap, where they lack domain knowledge required to safely use the tools they rely on for code generation. By integrating structured retrospectives, active learning assessments, and cognitive apprenticeship models, educators can promote comprehension oriented. CD offers a vital lens to ensure that next generation of software engineers possesses not just speed to generate code, but the depth to sustain it.

\begin{acks}
This work was supported by Helge Ax: son Johnsons Stiftelse, and Swedish Foundation for International Cooperation in Research and Higher Education (IB2020-8720). During the preparation of this work the author used DeepL to improve the language.  
\end{acks}

\bibliographystyle{ACM-Reference-Format}
\bibliography{references}

\end{document}